# Growth, Characterization and High Field Magneto-Conductivity of $Co_{0.1}Bi_2Se_3$ Topological Insulator

Rabia Sultana[1,2], Ganesh Gurjar[3], S. Patnaik[3] and V.P.S. Awana[1,2*]

[1] National Physical Laboratory (CSIR), Dr. K. S. Krishnan Road, New Delhi-110012, India

[2] Academy of Scientific and Innovative Research (AcSIR), NPL, New Delhi-110012, India

[3] School of Physical Sciences, Jawaharlal Nehru University, New Delhi-110067, India

**Abstract**

We report the crystal growth as well as transport properties of Co added $Bi_2Se_3$ ($Co_{0.1}Bi_2Se_3$) single crystals. The values of the lattice parameters a and b for Co added sample were observed to increase from a = b = 4.14(2) Å to 4.16(3) Å, whereas the c value decreased marginally from 28.70(3) Å to 28.69(5) as compared to the pure $Bi_2Se_3$. The Raman spectroscopy displayed higher Raman shift of corresponding $A_{1g}^1$, $E_g^2$ and $A_{1g}^2$ vibrational modes for $Co_{0.1}Bi_2Se_3$, and the resistivity curves with and without applied magnetic field shows a metallic behaviour. Both the crystals were subjected to magneto-resistance (MR) measurements under applied fields of 14Tesla. The value of MR is found to decrease from about 380% (5K, 14 Tesla) for $Bi_2Se_3$ to 200% for $Co_{0.1}Bi_2Se_3$. To elaborate the transport properties of pure and Co added $Bi_2Se_3$ crystals, the magneto-conductivity is fitted to the HLN (Hikami Larkin Nagaoka) equation and it is found that the charge conduction is mainly dominated by surface driven WAL (weak anti-localization) with negligible bulk WL (weak localization) contribution in both crystals alike. The MH curves of $Co_{0.1}Bi_2Se_3$ crystal at different temperatures displayed a combination of both ferromagnetic and diamagnetic behaviour. On the other hand, the Electron Paramagnetic Resonance (EPR) revealed that pure $Bi_2Se_3$ is diamagnetic whereas, Co orders ferro-magnetically with resonating field around 3422Oe at room temperature. The calculated value of Lande 'g' factor is around 2.04 ± 0.05. Summarily, the short letter discusses the interesting magneto-conductivity and complex magnetism of Co in $Co_{0.1}Bi_2Se_3$.

*Corresponding Author
Dr. V. P. S. Awana
E-mail: awana@nplindia.org
Ph. +91-11-45609357, Fax-+91-11-45609310
Homepage: awanavps.webs.c

**Introduction**

Recently, the discovery of a new class of materials "topological insulators (TIs)", has grabbed the attention of the entire condensed matter physics community owing to its growing practical applications importance, as well as in understanding of the exotic quantum states of matter [1-3]. TIs behave as a conventional insulator in its bulk/interior having insulating energy gaps and as conductor at the edges/surface (two dimensional (2D) / 3D materials) having gapless states [4-9]. These surface states are further protected by time reversal symmetry (TRS), coupled with spin momentum locking properties [4-9]. The presence of these unique topologically protected surface states are considered to be robust against backscattering from non magnetic impurities, defects or thermal agitations, which could be helpful to achieve low dissipation transport along with potential applications [1-10]. It is known that the surface robustness can be destroyed by doping the TIs with magnetic impurities by breaking of TRS, and opening a band gap near the Dirac point resulting into gapped surface states [11-14]. Also, doping three dimensional (3D) TIs ($Bi_2Te_3$, $Bi_2Se_3$, $Sb_2Te_3$) with transition metal elements (Cr, Fe, Mn, V, etc.) can bring about magnetic ordering, which may further enable to observe different exotic phenomena such as anomalous quantized Hall (AQH) state, magnetic monopole, topological magneto electric effect and the Faraday-Kerr effects [14-21]. In this regards, till date, several studies have been reported, which addresses the effect of magnetic ordering in various TIs viz., ferromagnetism (FM) has been observed in Cr doped $Bi_2Te_3$ and $Bi_2Se_3$ thin film [22-24], Fe doped $Bi_2Se_3$ nanoribbons [25,26], Mn and V doped $Bi_2Se_3$ thin films [27,28], Co doped $Bi_2Se_3$ [29-33], Mn and Ni doped $Bi_2Se_3$ [34-36], Fe doped $Bi_2Te_3$ [37], Mn doped bulk $Bi_2Te_3$ [38] and Fe and Co doped $Sb_2Te_3$ [39,40].

Besides, the realization of a number of exotic phenomena, magnetically doped TIs exhibits weak localization (WL) behaviour occurring due to the gapped surface states induced by the broken TRS [22,24,26,41]. Theoretically and experimentally it has been proposed that doping TIs with magnetic impurities results in quantum corrections to magneto-conductivity experiencing a crossover from weak anti-localization (WAL) effect to WL effect [22,24,41]. Accordingly, the presence of gapped surface states from the TRS breaking is responsible for this type of crossover (WAL to WL). Further, Cha et al showed the kondo effect along with the existence of both WAL and WL effects in $Bi_2Se_3$ nanoribbons [24, 25]. L. Bao et al. and M. Liu et al. showed crossover of quantum correction of magneto-conductivity from WAL to WL in Cr doped $Bi_2Te_3$ thin films and $Bi_2Se_3$ ultra thin films respectively [22, 24]. Consequently, the effects of magnetically doped TIs have aroused significant interest among various research groups [14-40]. Among the various discovered TIs ($Bi_{1-x}Sb_x$, $Bi_2Te_3$, $Bi_2Se_3$, $Sb_2Te_3$), $Bi_2Se_3$ is one of the most popular and widely studied 3D TI due to the presence of a single surface Dirac cone along with a large bulk band gap of 0.3eV, promising for room temperature applications [4, 15]. Although Co doping has been done earlier in $Bi_2Se_3$ [29-33], in the present study, we report the effect of Co doping in $Bi_2Se_3$ bulk single crystalline material and discuss mainly the magneto-conductivity (MC) data in terms of HLN model. It is concluded that though the value of MR at 5K is found to decrease to nearly half, the fitting to the HLN equation showed that the WAL dominates the MC curve with negligible bulk (WL) contribution partially coupled with the surfaces states in cases of pure and Co added crystals alike.

**Experimental Details**

The studied Co added $Bi_2Se_3$ ($Co_{0.1}Bi_2Se_3$) single crystal was grown by melting constituents elements using the self flux method via the conventional solid state reaction route, similar to that as reported by us for $Bi_2Se_3$ and $Bi_2Te_3$ single crystals [42,43]. Stoichiometric mixtures of high purity (99.999%) elements (cobalt, bismuth and selenium powders) were weighed accurately and grounded thoroughly inside an Argon filled glove box to obtain a homogeneous mixture. The well mixed powder was then pelletized into a rectangular form. Further, the rectangular pellet was sealed in an evacuated quartz tube ($10^{-3}$Torr) and kept inside an automated tube furnace. The quartz tube was heated at 950°C for 7.5 hours, hold for 24 hours and then slowly cooled (2°C/hour) to 650°C. Furthermore, the furnace was then switched off and the quartz tube was allowed to cool down naturally to room temperature. The detailed heat treatment is shown in the schematic diagram [Fig 1]. The synthesized sample obtained by breaking the quartz tube appeared shiny, which was mechanically cleaved for further measurements.

The structural characterization of the synthesized sample was carried out by employing Rigaku Miniflex II, Desktop X-ray Diffractometer (XRD) with Cu-Kα radiation (λ=1.5418 Å). The electrical and magnetic measurements were carried out in a Quantum Design Physical Property Measurement System (PPMS), Model 6000 using the standard four probe technique. Raman spectroscopy was used to analyze the vibrational modes of the studied $Co_{0.1}Bi_2Se_3$ single crystal using Renishaw Raman Spectrometer. Electron Paramagnetic Resonance (EPR) studies were performed at room temperature using a Bruker Biospin EMX10 (Model A300) spectrometer equipped with the X-band (f = 9.8 GHz) $TE_{011}$ resonance cavity.

**Results and Discussion**

The crystalline nature as well as the phase purity of as synthesized $Co_{0.1}Bi_2Se_3$ was determined from the XRD patterns. Figure 2 displays comparative XRD plots carried out on the freshly cleaved silvery surface of pure $Bi_2Se_3$ (Fig.2a) and Co added $Bi_2Se_3$ (Fig. 2b) crystals in the angular range of $2\theta_{min} = 10°$ to $2\theta_{max} = 80°$. The XRD data of pure $Bi_2Se_3$ sample has been taken from our previous work ref [42] in order to make a direct comparison between the two. The sharp (00l) reflections indicate the good crystalline quality of as synthesised Co added $Bi_2Se_3$. Additionally, the diffraction peaks of both pure and Co added $Bi_2Se_3$ samples are seen to be aligned along (00l) plane and are well indexed. Accordingly, XRD pattern clearly reveals that the synthesized $Co_{0.1}Bi_2Se_3$ sample is single crystalline in nature and oriented in c-axis, similar to the pure $Bi_2Se_3$ sample [42].

Figure 3 represents the powder XRD pattern of as synthesized $Co_{0.1}Bi_2Se_3$ crystal. This time the crystal is gently crushed before taking the XRD pattern. Full Prof Suite Toolbar Software was employed to perform the Rietveld fitting of the obtained powder XRD (PXRD) pattern. The PXRD pattern showed nearly no impurity phase of any kind within the XRD limits. One can assume that the studied $Co_{0.1}Bi_2Se_3$ crystal is of reasonably good quality. The Rietveld fitting of the PXRD shows that the as synthesized $Co_{0.1}Bi_2Se_3$ is crystallized in rhombohedral structure with R-3m space group. The refined lattice parameters for the resultant $Co_{0.1}Bi_2Se_3$ crystal are found to be a=b= 4.16(3) Å and c = 28.69(5) Å respectively. The values of the lattice parameters a and b were observed to increase from a = b = 4.14(2) Å for pure $Bi_2Se_3$ to 4.16(3) Å for Co added $Bi_2Se_3$. Further, a slight decrease in the c- axis lattice parameter from 28.70 (3) Å for pure $Bi_2Se_3$ to 28.69(5) for Co added $Bi_2Se_3$ was observed, which is in agreement to the earlier reported data [29-33]. Both the increment and decrement in the refined lattice constants are opposite to those observed in topological superconductors (TSCs). In TSCs such as Cu doped $Bi_2Se_3$ and Sr intercalated $Bi_2Se_3$ the values of the lattice constant, c was found to increase, whereas the values of a and b appears to be the same as the parent compound i.e., $Bi_2Se_3$ [42, 44, 45]. It seems in case of TSCs (Cu doped $Bi_2Se_3$), the Cu enters $Bi_2Se_3$ either as an inter-calant in the van der Waals (vdW) gap or acts as a substitution type defect to replace Bi, thus increasing c parameter but leaving a, and b unaffected [30]. On the other hand, the magnetic dopants viz. Co, Mn, Cr etc., the c-parameter decrease and a, b slightly increases. As reported, a small amount of Co atoms may occupy the Bi lattice site, whereas the remaining Co atoms most likely form interstitial atoms or ions within the layers of the crystal lattice by entering the vdW gap of the crystal structure [30]. The decrease in c parameter in the magnetically doped TIs occurs due to the smaller atomic radius of the magnetic dopants viz. Co, Mn, Cr etc., in comparison to Bi. In our case, the magnetic dopant Co exhibits smaller atomic radius (1.25 Å) than that of Bi (1.56 Å) [30, 32]. In any case, because the present study is about added Co i.e., $Co_xBi_2Se_3$ with x = 0.10, it may be advisable to focus further on $Bi_{2-x}Co_xSe_3$ with higher x values as well.

To understand the impact of Co doping on the phonon dynamics of $Bi_2Se_3$ single crystal, we performed Raman spectroscopy measurements at room temperature of both ($Bi_2Se_3$ and $Co_{0.1}Bi_2Se_3$) samples, followed by direct comparison between the two. The vibrational modes at room temperature of both pure ($Bi_2Se_3$) and Co added $Bi_2Se_3$ ($Co_{0.1}Bi_2Se_3$) were observed using spectral resolution of $0.5 cm^{-1}$ and 2400 I/mm grating, having an excitation photon energy of 2.4eV (514nm). The laser power applied at the surface was kept less than 3mW to avoid the sample damage. The representing three main distinct Raman active vibrational peaks namely $A_{1g}^1$, $E_g^2$ and $A_{1g}^2$ are seen clearly in the measured spectral range of $10–300 cm^{-1}$ for both pure $Bi_2Se_3$ (Fig.4a) and Co added $Bi_2Se_3$ (Fig. 4b). Namely, the obtained values for the vibrational modes corresponding to $A_{1g}^1$, $E_g^2$ and $A_{1g}^2$ are 72, 131 and $174.5 cm^{-1}$ for $Bi_2Se_3$ (Fig. 4a) and 73, 132 and $175.5 cm^{-1}$ respectively for $Co_{0.1}Bi_2Se_3$ (Fig. 4b). The Raman shift of the corresponding $A_{1g}^1$, $E_g^2$ and $A_{1g}^2$ vibrational modes is higher for Co added $Bi_2Se_3$ ($Co_{0.1}Bi_2Se_3$) sample as compared to the pure ($Bi_2Se_3$) sample (Fig. 4c). Consequently, the Raman shift of the vibrational modes increases with Co doping, where as the peak intensities were observed to be decreasing simultaneously. Further, the experimentally observed vibrational modes values for the Co added sample exhibited higher values of Raman shift in comparison to the other well known TIs ($Bi_2Te_3$, $Sb_2Te_3$) as reported by some of us recently [46]. Correspondingly, the existence of higher Raman shifts values in case of $Co_{0.1}Bi_2Se_3$ may possibly occur due to the stronger bonding forces in comparison to $Bi_2Te_3$ and $Sb_2Te_3$ TIs, along with lighter atomic weight of Co as compared to Bi, Se, Te and Sb atom respectively [46,47].

Figure 5 displays the temperature dependent electrical resistivity measurements of the as synthesized $Co_{0.1}Bi_2Se_3$ in the absence of magnetic field (0Tesla). As the temperature increases from 5K to 300K the resistivity is observed to increase simultaneously. Consequently, the resistivity curve exhibits a metallic behaviour up to 300K. To inter compare the resistivity behaviour of $Co_{0.1}Bi_2Se_3$ with that of pristine $Bi_2Se_3$, inset of figure 5, shows the normalised ($\rho/\rho_{250K}$) resistivity as a function of temperature from 250K down to 5K. It is clear that though both Co added and pristine $Bi_2Se_3$ are metallic in nature, the relative conductivity is less for the Co added one. This is envisaged from the fact that the normalised resistivity slope is more for pure crystal than the Co added one.

Figure 6 shows the temperature dependent electrical resistivity measurements under different applied magnetic fields. The temperature ranges from 2K to 100K and the field varies from 0Tesla to 7Tesla. The values of resistivity curves are observed to increase with increase in temperature and yet depicting metallic behaviour in all applied fields. Similar, to the parent single crystal i.e., $Bi_2Se_3$, the $\rho(T)H$ curves shows a kind of saturation at lower temperature, say below 20K, which is related to Fermi liquid behaviour [48]. As far as relative impact of magnetic field on resistivity in comparison to pure $Bi_2Se_3$ is concerned, the same is seemingly less in Co added crystal. This will be more visible, when we show the isothermal magneto-resistance (RH) at a fixed temperature, say 5K in next section.

Figure 7 depicts the percentage change of magneto resistance (MR %) under different applied magnetic fields (H ⊥ c-axis and H // c-axis) of up to 14Tesla at 5K for both pure ($Bi_2Se_3$) and Co added $Bi_2Se_3$ ($Co_{0.1}Bi_2Se_3$) single crystals. The magneto-transport measurements are taken with same geometry as in our own previous work on $Bi_2Te_3$ single crystal [49]. Principally, the magnetic field is applied perpendicular to the transport current; the schematic diagram is shown in inset of Fig.7 as well. The value MR is calculated using the MR (%) = {[$\rho(H)$ - $\rho(0)$] / $\rho(0)$}*100, where $\rho(H)$ and $\rho(0)$ are the resistivity with and without applied magnetic field (H) respectively. At 5K, when the applied magnetic field of up to 14Tesla is perpendicular to the c axis, $Bi_2Se_3$ exhibits the larger, linear positive MR value (~380%), in comparison to the Co added $Bi_2Se_3$ ($Co_{0.1}Bi_2Se_3$) sample (~200%) at 5K. However, when the applied magnetic field is changed from perpendicular to parallel (H // c-axis), $Bi_2Se_3$ displays greater MR value (~30%) in comparison to the Co added $Bi_2Se_3$ ($Co_{0.1}Bi_2Se_3$) sample (~10%). Consequently, we can say that doping of magnetic impurity (Co) in $Bi_2Se_3$ single crystal results in lower MR value. It is well known by now that non-saturating high linear MR is intrinsic property of TRS protected TIs [1-14]. The decrease in value of linear MR from about 380% to 200% at 5K and 14Tesla field for $Co_{0.1}Bi_2Se_3$, in comparison to pristine $Bi_2Se_3$ shows that TRS is affected by magnetic Co doping. Although the possible impact of Co doping (decrease in MR) may remain the same, it is advisable that the MR measurements be performed at various other temperature including lower (2K) and higher 10, 20K etc. Further lowering the temperature to mK range may eventually even show the AQH.

To elaborate more on the impact of magnetic Co doping on TRS we show the change of magneto-conductivity (MC) curves of both pure ($Bi_2Se_3$) and Co added $Bi_2Se_3$ ($Co_{0.1}Bi_2Se_3$) single crystals in Fig. 8. Here, the low field MC has been described using the Hikami - Larkin - Nagaoka (HLN) theory [50]:

$$\Delta\sigma(H) = \sigma(H) - \sigma(0)$$

$$= -\frac{\alpha\, e^2}{\pi h}\left[\ln\left(\frac{B_\varphi}{H}\right) - \Psi\left(\frac{1}{2} + \frac{B_\varphi}{H}\right)\right]$$

Where, $\Delta\sigma(H)$ represents the change of magneto-conductivity, α is a coefficient signifying the overall strength of the WAL, e denotes the electronic charge, h represents the Planck's constant, Ψ is the digamma function, H is the applied magnetic field, $B_\varphi = \frac{h}{8e\pi H l_\varphi}$ is the characteristic magnetic field and $l_\varphi$ is the phase coherence length. The coefficient α, determining the type of localization as well as the number of coherent transport channels should have values -0.5 for single surface conducting channel and between -0.5 to -1.5 for multi parallel conduction channels [51-57]. The value of α is positive for WL and negative for WAL. Also, a large negative value can be caused by WAL in the bulk and two decoupled surface states, each contributing with α = -0.5. The experimentally fitted value of α varies widely, due to the problems arising from differentiating the bulk and surface contributions clearly. Also, the HLN fitting becomes challenging due to the competition between WAL and WL and hence, the extraction of α value becomes difficult. As reported, α may lie between –0.4 and –1.1, for single surface state, two surface states, or intermixing between the surface and bulk states [54, 57].

The inset of Figure 8 shows the MC curves of both pure ($Bi_2Se_3$) and Co added $Bi_2Se_3$ ($Co_{0.1}Bi_2Se_3$) single crystals fitted using HLN equation at 5K under lower applied magnetic fields of ± 0.5Tesla. The MC data for both pure ($Bi_2Se_3$) and Co added $Bi_2Se_3$ ($Co_{0.1}Bi_2Se_3$) single crystals was well fitted using HLN equation at 5K as shown by blue and red curves (solid lines) respectively. Here, α and $l_\varphi$ denotes the fitting parameters. The value of α at 5K for the pure ($Bi_2Se_3$) sample, extracted using the HLN equation is ~ - 0.96 ± 0.017 whereas, the $l_\varphi$ obtained is 10.31nm. The value of α slightly differs from -1, signifying the existence of two decoupled surface states with an additional bulk channel partially coupled with the surface states. This means WAL dominates the MC curve with negligible WL component. In other words, the observed WAL is not purely due to the surface states, but rather surface state dominated transport with negligible bulk contribution. The value of α = ~ - 0.96 ± 0.017 for the pure ($Bi_2Se_3$) sample are in agreement with the reported values of WAL and WL originated 2D surface states of TIs [22, 24]. For the Co added sample ($Co_{0.1}Bi_2Se_3$) the α value is ~ -0.93 ± 0.003 and $l_\varphi$ is 11.27nm respectively, indicating that again WAL dominates the MC curve with negligible WL contribution to the conduction mechanism.

However, in case of Co added bulk $Bi_2Se_3$ crystal although, the contribution of WL is negligible but is more in comparison to the pure sample. This is because the value of α (~ - 0.93 ± 0.003) for Co added bulk $Bi_2Se_3$ crystal deviates further more from -1 in comparison to the pure sample α (~ - 0.96 ± 0.017) value. The values of α between as ~ - 0.96 ± 0.017 and ~ -0.93 ± 0.003 for pure and Co added bulk $Bi_2Se_3$ crystal indicates that WAL contribution is maximized, whereas the WL contribution is minimum to the overall conduction mechanism. It is known, that the mechanism of conduction is quite challenging when both WAL (surface states) and WL (bulk states) competes with each other [22, 24] Further, as reported, the values of α and $l_\varphi$ may differ considerably from sample to sample (thin films, single crystals, nano wires) depending on the nature of localizations taking place i.e., WAL or WL or exhibiting relative contribution from both WAL and WL effects [22]. In our case the pristine sample exhibits α value as ~ - 0.96 ± 0.017 and Co added as ~ -0.93 ± 0.003, suggesting that the observed WAL effect may result mainly from the two decoupled surface states (WAL) with small contribution from WL states of the bulk channels.

Figure 9 (a) displays the magnetization (M) as a function of applied magnetic field (H) measurements at different temperatures (2K, 50K, 100K and 300K) for the studied Co added $Bi_2Se_3$ ($Co_{0.1}Bi_2Se_3$) single crystal. The magnetic field applied is up to 4Tesla. Broadly, the M-H curve seems to be diamagnetic at 300K and then changes from diamagnetic to paramagnetic as the temperature is decreased from 300K to 50K. Further, as the temperature is lowered to 2K, the M-H curve looks like ferromagnetic with clearly superimposed diamagnetic component (marked with straight black lines) at higher field above 2Tesla. Although, the nature of the graph is in agreement with the earlier reported literature [29, 40], the same needs further analysis. As reported earlier, pure $Bi_2Se_3$ sample exhibits a diamagnetic behaviour [29, 30]. In doped samples, Cobalt is believed to exist both in +2 and +3 states and the exchange interaction between these two states is claimed to be anti-ferromagnetic [29, 40]. Consequently, one can observe weak anti-ferromagnetic ordering with diamagnetic background which occurs due to the parent $Bi_2Se_3$. However, before accepting the explanation given above, one must separate the pure $Bi_2Se_3$ diamagnetic contribution from the overall magnetization. In our case, in fact a combination of both diamagnetic and ferromagnetic behaviour is evident, which is discussed in the next paragraph.

Further, to verify the magnetic phase and magnetic ordering of Co in $Co_{0.1}Bi_2Se_3$, we present the magnetization versus applied magnetic field curves [fig. 9(b)] plotted using the formula

$$M_{cobalt} = M_{total} - \chi_d * H$$

Where, $M_{cobalt}$ and $M_{total}$ represents the Co and total magnetization in emu/g, $\chi_d$ denotes the diamagnetic susceptibility and H is the applied magnetic field. The diamagnetic susceptibility at all temperatures is obtained by linearly fitting the M-H curves [fig. 9(a)] at higher fields. The magnetization of Cobalt ($M_{cobalt}$) was thus obtained by subtracting the diamagnetic signals from the total magnetization. Accordingly, the resultant magnetization of $Co_{0.1}Bi_2Se_3$ was observed as a combination of both ferromagnetic and diamagnetic signals. These results are shown in fig. 9(b). It is also clear that though at 2K, the saturation of Co moments occur at nearly above 2Tesla, the coercive field is negligible. The anisotropic magnetization measurements i.e., applying magnetic field in both parallel and perpendicular directions to the crystal planes could provide more information. This is important because in-plane and out of plane magnetization of pure and in particular the magnetically doped TIs do vary. At present due to lack of PPMS sample rotator, we could not perform this study. However from our magnetization results it is clear that Co orders ferro-magnetically (FM) right from room temperature down to 2K, with saturation moment of around 0.2emu/g above 2Tesla and the coercive field seen is negligible. This is the situation when field is applied perpendicular to the plane of the crystal. To elaborate more on magnetism of Co, in next section we discuss the EPR of our studied crystals.

Till date, the investigation into magnetism in transition metal-doped TIs is still in the initial stage and seems contradictory. Accordingly, an in depth study for the complete understanding of the surface as well as bulk magnetism separately is required. In this regard, we performed the room temperature EPR measurement of both pure and Co added $Bi_2Se_3$ ($Co_{0.1}Bi_2Se_3$) sample using Bruker Biospin EMX10 (Model A300) spectrometer. The EPR is a powerful, versatile, non-destructive, and non intrusive analytical technique based on Zeeman splitting to detect and characterize the transition metal elements in classical semiconductors as well as in TIs. Here, the EPR spectrum for both pure and Co added $Bi_2Se_3$ ($Co_{0.1}Bi_2Se_3$) samples were collected at a frequency of 9.8GHz and at lower magnetic fields reaching up to 12000Oe. Figure 10 displays the EPR spectrum for both pure and Co added $Bi_2Se_3$ single crystals. No spectrum was observed for the $Bi_2Se_3$ single crystal, suggesting that the same is diamagnetic in nature. On the other hand, the Co added $Bi_2Se_3$ ($Co_{0.1}Bi_2Se_3$) sample showed broad spectrum exhibiting ferromagnetism behaviour as observed from fig.10. Further, we calculated the Lande factor (g factor) of $Co_{0.1}Bi_2Se_3$ sample using the following expression:

$$g = h\upsilon / \mu_B H_r$$

where, h is the Planck's constant, $\upsilon$ is the frequency, $\mu_B$ is the Bohr magneton and $H_r$ is the field resonance. The field resonance ($H_r$) value is taken from the EPR spectrum as 3422Oe. Accordingly, the calculated Lande (g) factor for $Co_{0.1}Bi_2Se_3$ was obtained to be 2.04± 0.05. The g value (2.04± 0.05) slightly differs from the free electron (~2) of un-clustered Co, due to the instrumental error. It is clear from fig.10 that Co orders ferromagnetically in $Co_{0.1}Bi_2Se_3$ and the raw magnetization (MH) results (present here – fig. 9a and in ref. 29) may not be the true representative as the same are seemingly dominated by the $Bi_2Se_3$ diamagnetic back ground. This is clear after due analysis (separating the $Bi_2Se_3$ diamagnetic part) shown in fig. 9(b) along with the EPR results (fig. 10) that Co orders FM in $Co_{0.1}Bi_2Se_3$.

**Conclusion**

Summarily, we have successfully grown Co added $Bi_2Se_3$ ($Co_{0.1}Bi_2Se_3$) single crystal via the self flux melt grown method and discussed their structural, spectroscopic (Raman), high field (14Tesla) magneto-transport and magnetization characteristics. Although, the MR obtained for Co added sample (200%, 5K, 14Tesla) was much less (nearly half) than the pure sample, the MC being fitted to the HLN equation showed that the charge conduction mechanism is WAL dominating with negligible bulk (WL) contribution, partially coupled with the surfaces states in cases of pure and Co added crystals alike. The MH curves of $Co_{0.1}Bi_2Se_3$ crystal at different temperatures displayed a combination of both ferromagnetic and diamagnetic behaviour. The EPR measurements revealed diamagnetic and ferromagnetic nature of pure and Co added $Bi_2Se_3$ ($Co_{0.1}Bi_2Se_3$) crystals respectively.

**Acknowledgements**

The authors from CSIR-NPL would like to thank their Director NPL, India, for his keen interest in the present work. Authors further thank Mrs. Shaveta Sharma for Raman studies and Saurabh Pathak for EPR studies. S. Patnaik thanks DST-SERB project (EMR/2016/003998) for the low temperature high magnetic facility at JNU, New Delhi. Rabia Sultana and Ganesh Gurjar thank CSIR, India, for research fellowship. Rabia Sultana thanks AcSIR-NPL for Ph.D. registration.

**Figure Captions**

**Figure 1:** Schematic heat treatment diagram for $Co_{0.1}Bi_2Se_3$ single crystal.

**Figure 2:** X-ray diffraction pattern of as synthesized (a) $Bi_2Se_3$ and (b) $Co_{0.1}Bi_2Se_3$ single crystal.

**Figure 3:** Rietveld fitted room temperature XRD pattern for powder $Co_{0.1}Bi_2Se_3$ single crystal.

**Figure 4:** Raman spectra of (a) $Bi_2Se_3$ (b) $Co_{0.1}Bi_2Se_3$ (c) both $Bi_2Se_3$ and $Co_{0.1}Bi_2Se_3$ single crystal.

**Figure 5:** Temperature dependent electrical resistivity of $Co_{0.1}Bi_2Se_3$ single crystal in a temperature range of 300K to 5K. Inset shows temperature dependent normalized resistivity curve at different temperatures for both $Bi_2Se_3$ and $Co_{0.1}Bi_2Se_3$ single crystal.

**Figure 6:** Temperature dependent electrical resistivity under different applied magnetic field for $Co_{0.1}Bi_2Se_3$ single crystal.

**Figure 7:** MR (%) as a function of magnetic field (H) perpendicular and parallel to ab plane at 5K for both $Bi_2Se_3$ and $Co_{0.1}Bi_2Se_3$ single crystal, inset shows schematic diagram for magneto resistivity measurements.

**Figure 8:** WAL related magneto-conductivity for both $Bi_2Se_3$ and $Co_{0.1}Bi_2Se_3$ single crystal at 5K fitted using the HLN equation. Inset shows the zoom in view of the fitted curve at low fields up to ± 0.5Tesla.

**Figure 9 (a):** Magnetization as a function of applied magnetic for $Co_{0.1}Bi_2Se_3$ single crystal at different temperatures (2K, 50K, 100K and 300K).

**Figure 9 (b):** M - H curves of $Co_{0.1}Bi_2Se_3$ single crystal at different temperatures (2K, 50K, 100K and 300K) showing two contributions: diamagnetic as well as ferromagnetic.

**Figure 10:** EPR spectroscopy of both $Bi_2Se_3$ and $Co_{0.1}Bi_2Se_3$ single crystal recorded at room temperature.

Fig. 1

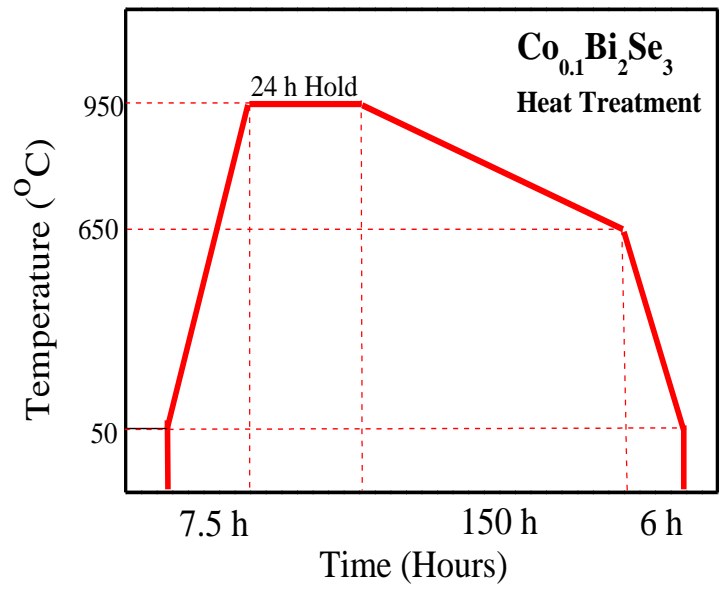

Fig. 2

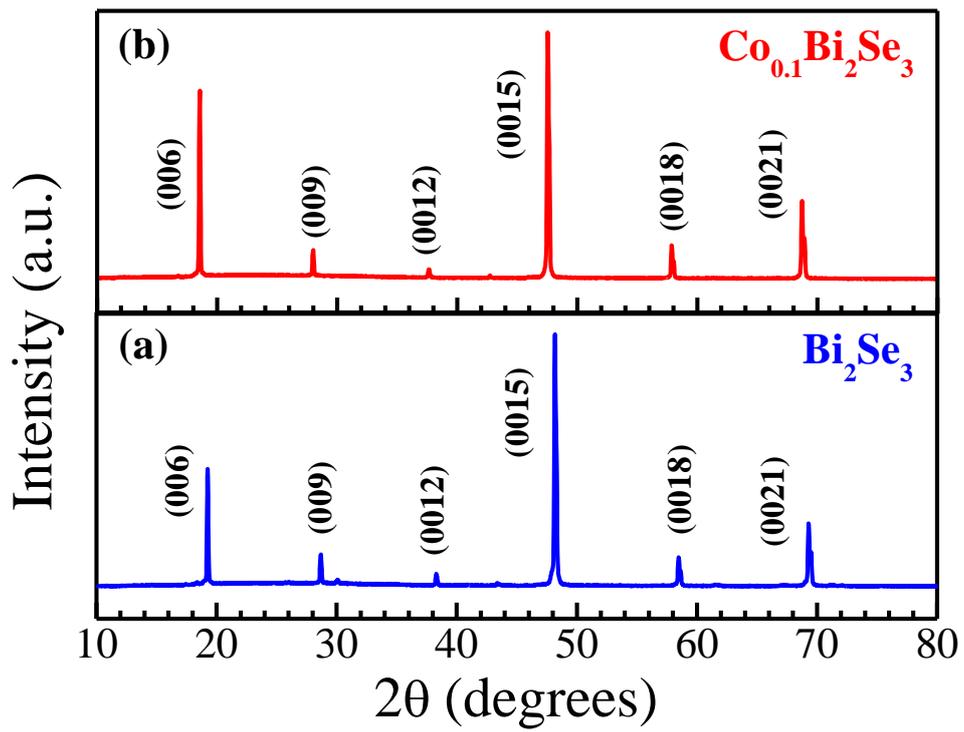

Fig. 3

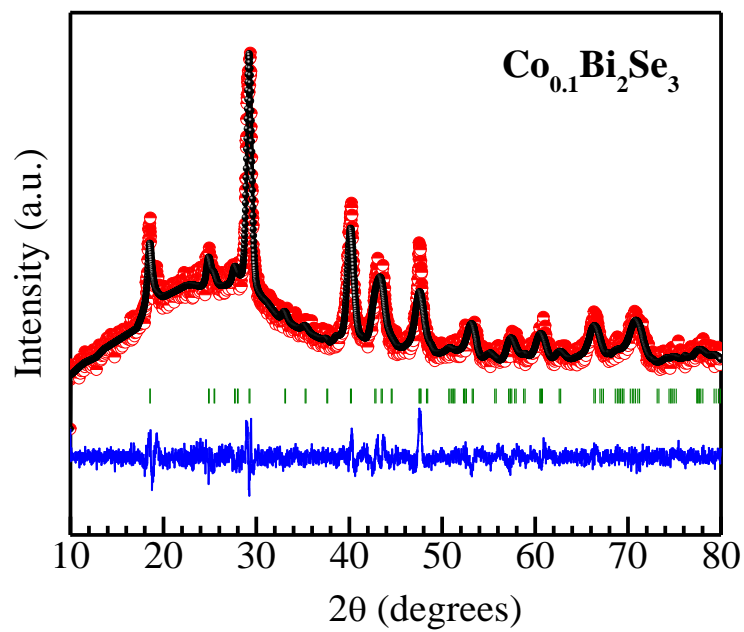

Fig. 4

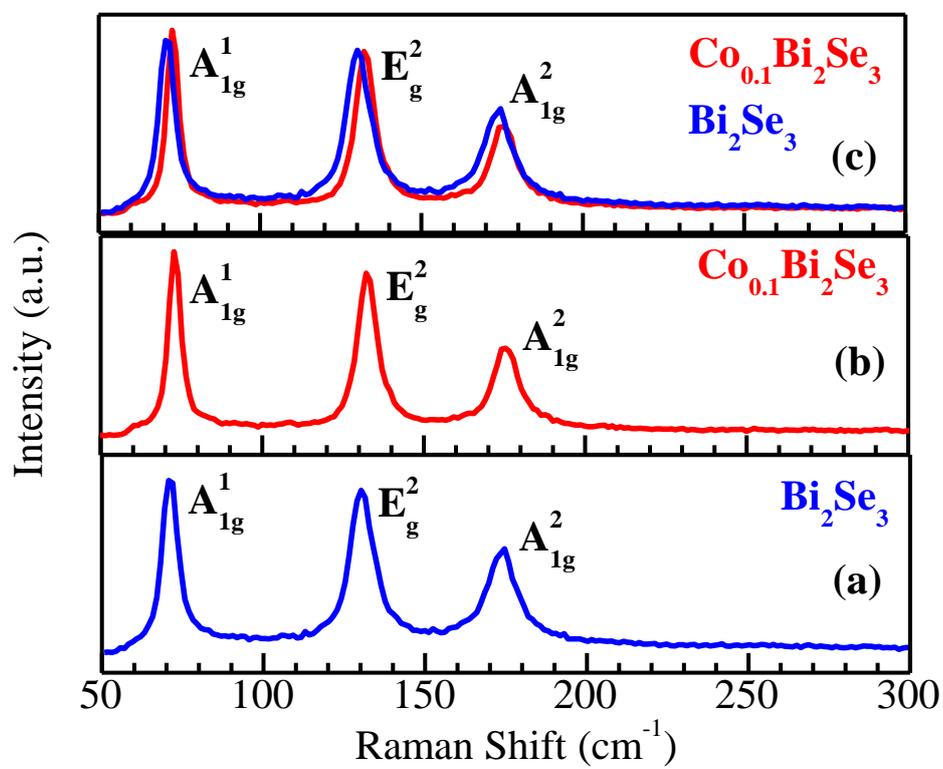

Fig. 5

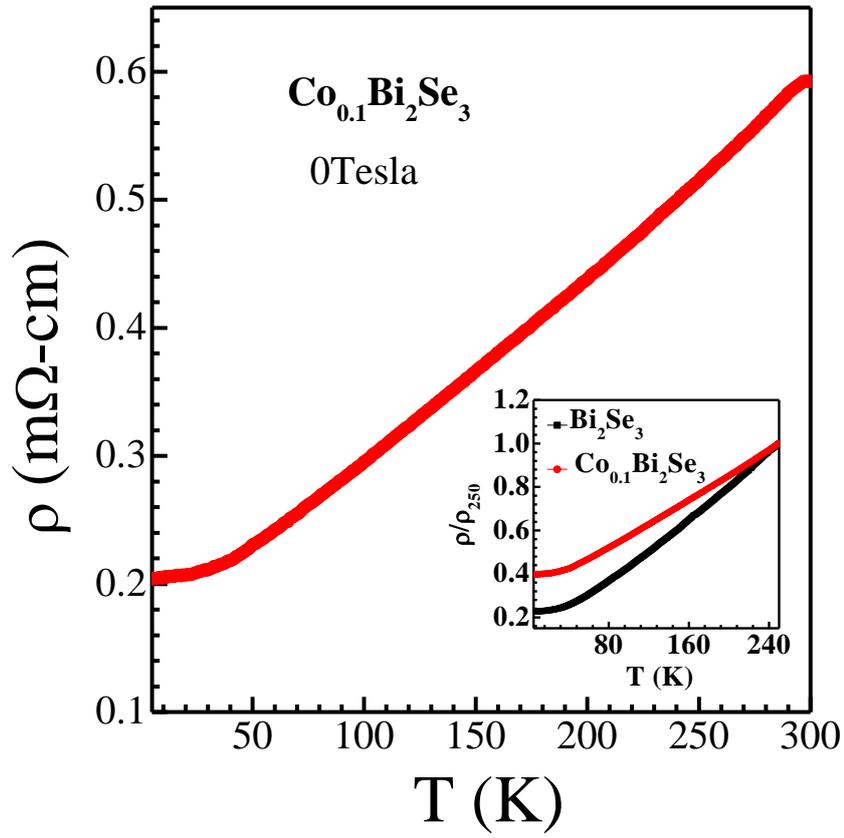

Fig. 6

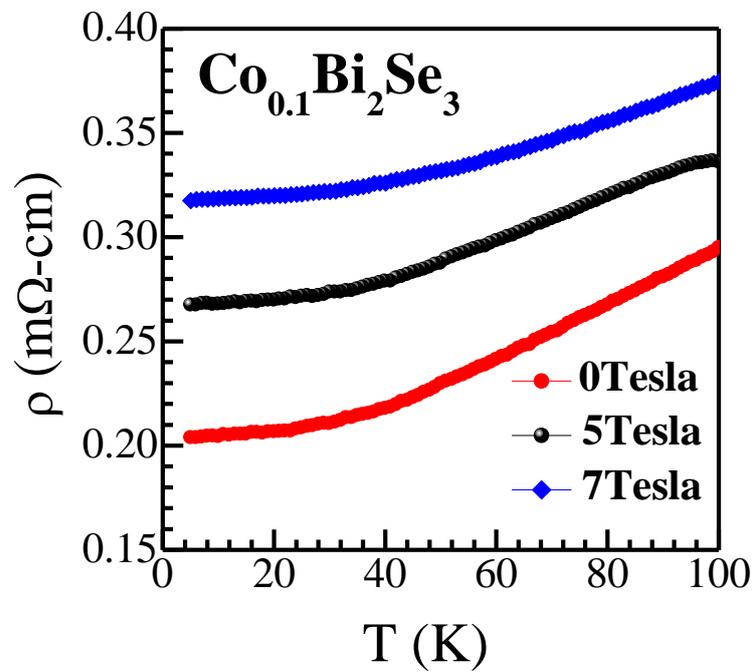

Fig. 7

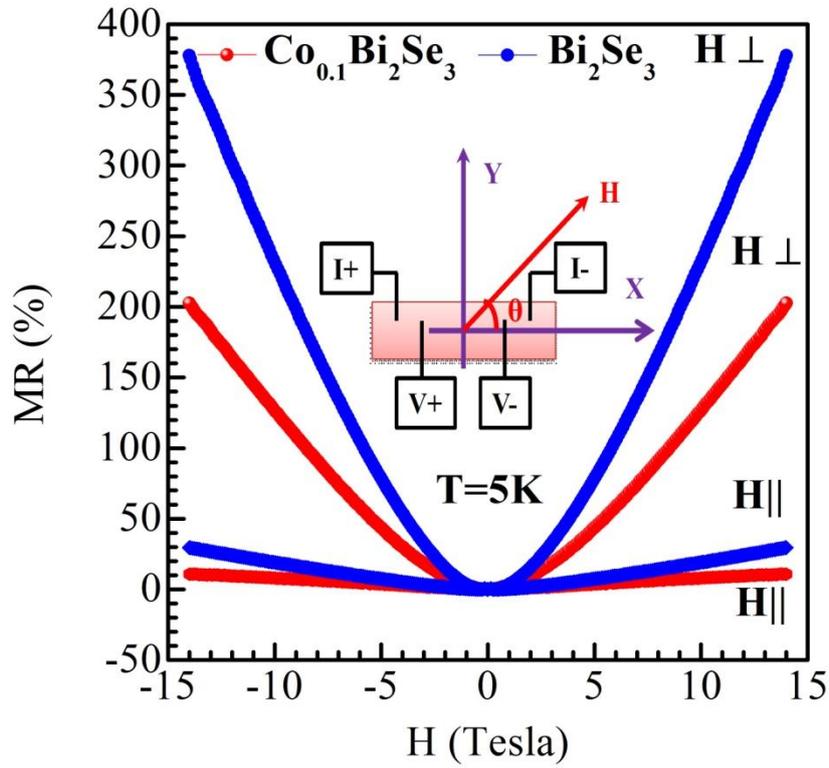

Fig. 8

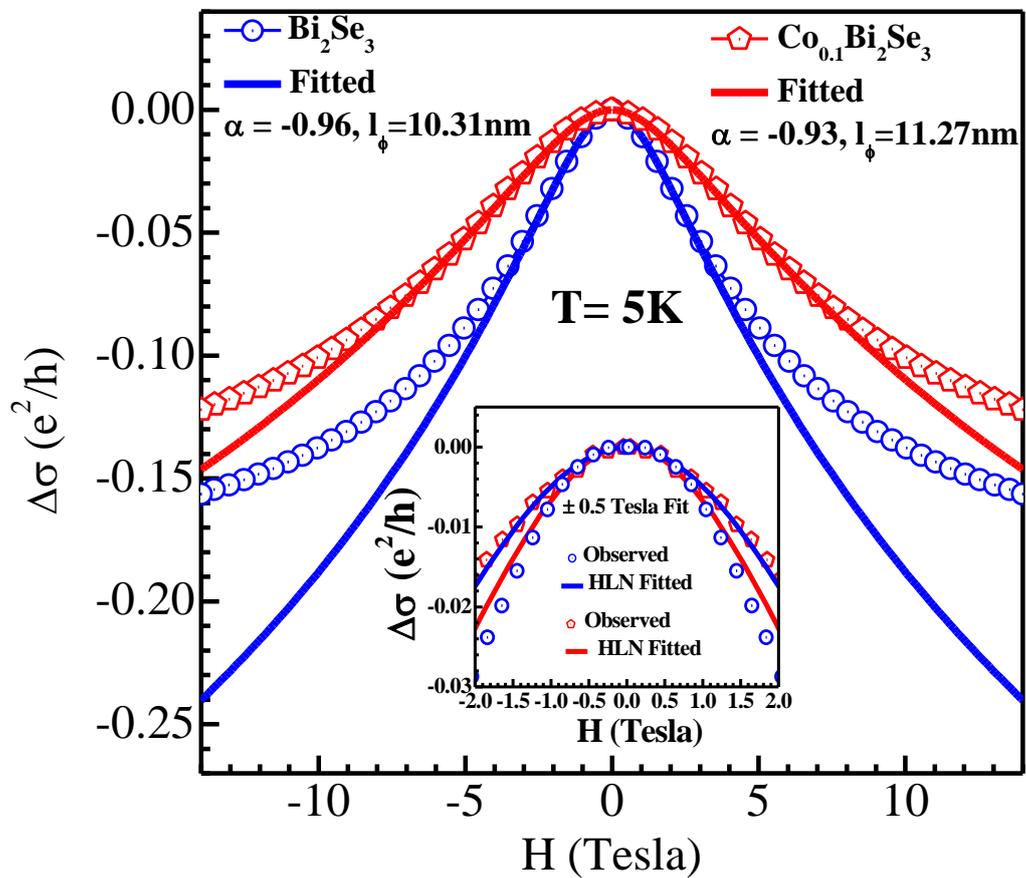

Fig. 9(a)

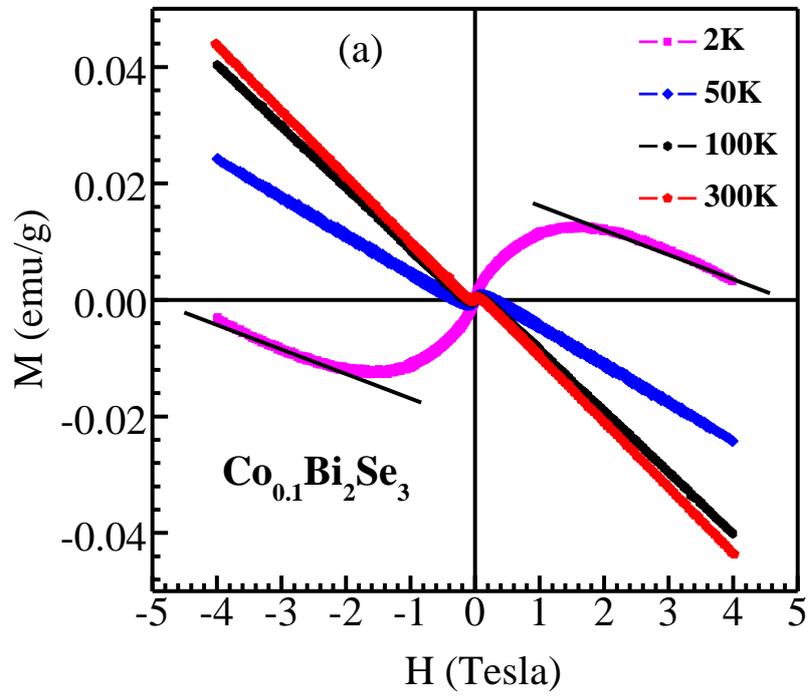

Fig. 9(b)

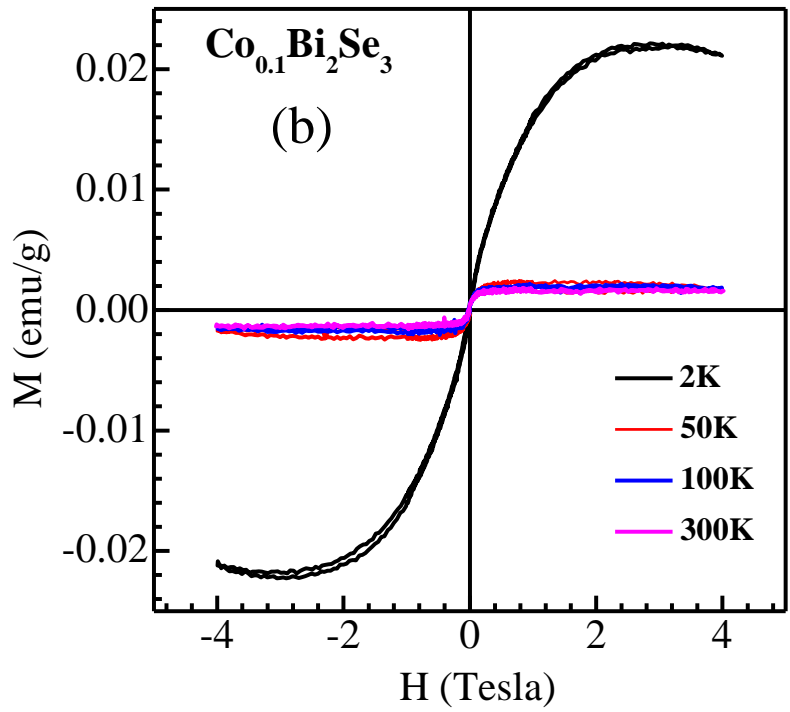

Fig. 10

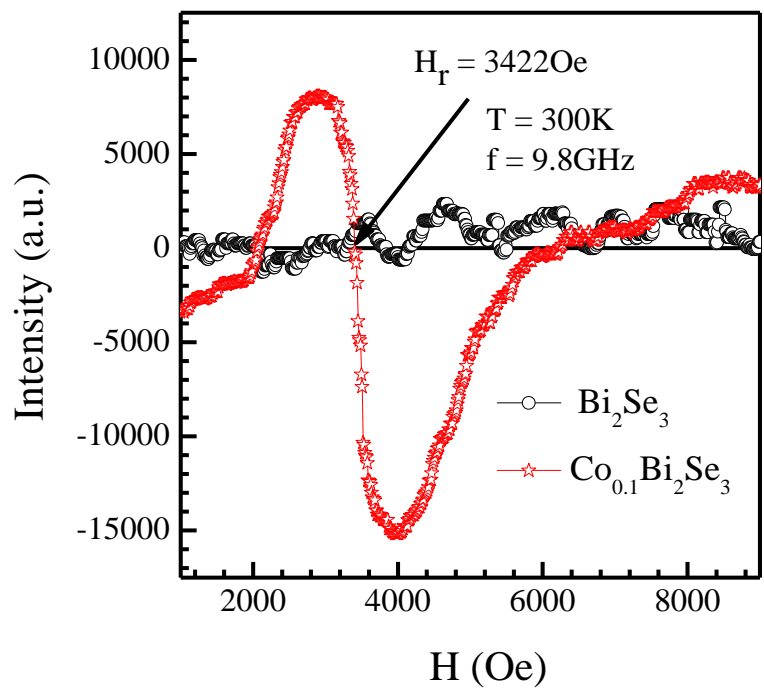